\documentclass[aps,prb,twocolumn,amssymb,showpacs,amsmath,superscriptaddress,floatfix]{revtex4}
\usepackage{graphicx}
\usepackage{bm}

\begin{document}
\title {Local breaking of the spin degeneracy in the vortex states of Ising superconductors:
Induced antiphase ferromagnetic order}

\author{Hong-Min Jiang}
\affiliation{School of Science, Zhejiang University of Science and
Technology, Hangzhou 310023, China}
\author{Xiao-Yin Pan}
\affiliation{Department of Physics, Ningbo University, Ningbo
315211,China}

\date{\today}

\begin{abstract}
Ising spin-orbital coupling is usually easy to identify in the Ising
superconductors via an in-plane critical field enhancement, but we
show that the Ising spin-orbital coupling also manifests in the
vortex physics for perpendicular magnetic fields. By
self-consistently solving the Bogoliubov-de Gennes equations of a
model Hamiltonian built on the honeycomb lattice with the Ising
spin-orbital coupling pertinent to the transition metal
dichalcogenides, we numerically investigate the local breaking of
the spin and sublattice degeneracies in the presence of a
perpendicular magnetic field. It is revealed that the ferromagnetic
orders are induced inside the vortex core region by the Ising
spin-orbital coupling. The induced magnetic orders are antiphase in
terms of their opposite polarizations inside the two
nearest-neighbor vortices with one of the two polarizations coming
dominantly from one sublattice sites, implying the local breaking of
the spin and sublattice degeneracies. The finite-energy peaks of the
local-density-of-states for spin-up and spin-down in-gap states are
split and shifted oppositely by the Ising spin-orbital coupling, and
the relative shifts of them on sublattices $A$ and $B$ are also of
opposite algebraic sign. The calculated results and the proposed
scenario may not only serve as experimental signatures for
identifying the Ising spin-orbital coupling in the Ising
superconductors, but also be prospective in manipulation of electron
spins in motion through the orbital effect in the superconducting
vortex states.
\end{abstract}

 \maketitle

\section{introduction}
The superconductivity uncovered in atomically thin two-dimensional
(2D) forms of layered transition metal dichalcogenides (TMDs) have
recently attracted remarkable scientific and technical
interests~\cite{Bromley1,Boker1,Zhu1,Xiao1,JTYe1,Taniguchi1,Kormanyos1,Zahid1,Cappelluti1,XXi1,WShi1,jmlu1,Saito1,XXi2}.
Although these superconductors belong to the conventional $s$-wave
superconductivity with low transition
temperature~\cite{Bromley1,Boker1,Zhu1,Xiao1,JTYe1,Taniguchi1,Kormanyos1,Zahid1,Cappelluti1,XXi1,WShi1,jmlu1,Saito1,XXi2},
the uniqueness of the TMDs makes them alluring to the researchers.
On one hand, similar to graphene, these materials have a honeycomb
lattice structure, and exhibit a valley degree of freedom with
minima/maxima of conduction/valence bands at the corners
$\mathbf{K}$ and $-\mathbf{K}$ of the Brillouin zone. On the other
hand, unlike graphene, the in-plane mirror symmetry is broken in the
TMDs, leading to a strong atomic Ising type spin-orbital coupling
(ISOC)~\cite{Zhu1,Xiao1,Kormanyos1,Zahid1,Cappelluti1}. The ISOC
strongly pins the electron spins to the out-of-plane directions and
have opposite directions in opposite valleys ($\mathbf{K}$ and
$-\mathbf{K}$)~\cite{Zhu1,Xiao1,Kormanyos1,Zahid1,Cappelluti1,jmlu1,XXi2,btzhou1,Sharma1},
so that it preserves time-reversal symmetry and is compatible with
superconductivity. Due to the strong pinning of electron spins in
the out-of-plane directions, external in-plane magnetic fields are
much less effective in aligning electron spins, and lead to the
in-plane upper critical field $H_{c2}$ of the system several times
larger than the Pauli limit~\cite{XXi1,jmlu1}.

Nevertheless, an out-of-plane magnetic field will generate the
magnetic flux in conductors due to the dominating orbital effect
over the Zeeman splitting. It is well known that the superconductors
expel the magnetic flux from their interior, the so called Meissner
effect. While some superconductors expel the magnetic field globally
(they are called type I superconductors), a type II superconductor
will only keep the whole magnetic field out until a first critical
field $H_{c1}$ is reached. Then vortices start to appear. A vortex
is a local magnetic flux quantum that penetrates the superconductor,
where the superconducting (SC) order parameter drops to zero to save
the rest of the SC state in metal from being destroyed. While the
ISOC exemplifies itself as the spin-valley locking in the momentum
space, it acts as coupling between spins and the orbital derived
effectively periodic spin and sublattice dependent fluxes in real
space with the quantization axis along the out-of-plane direction.
This is to say the spins, sublattices and the effectively periodic
fluxes are bound together by the ISOC in real space. Thus, the local
breaking of the spin and sublattice degeneracies may be expected if
the fluxes are altered locally, and the spin orders in real space
may also be expected to emerge.

In this paper, we numerically demonstrate that the spin and
sublattice degeneracies break locally with an induced ferromagnetic
order inside the vortex core of the Ising superconductors, as a
result of the contrasting variation of the effectively periodic
fluxes for sublattices $A$ and $B$ caused by the out-of-plane
magnetic field. By self-consistently solving the Bogoliubov-de
Gennes (BdG) equations of the Hamiltonian, it is shown that there is
no magnetic order induced inside the vortex core when the ISOC is
zero. Accordingly, the curves of the local-density-of-states (LDOS)
for the spin-up and spin-down in-gap states are almost identical,
forming a series of discrete energy peaks inside the core region.
The inclusion of the ISOC induces a ferromagnetic order inside the
vortex core, where the SC order parameter is suppressed. The induced
magnetic orders are antiphase in terms of their opposite
polarizations inside two nearest-neighbor (NN) vortices with one of
the two polarizations coming dominantly from one sublattice sites.
The finite-energy peaks of the LDOS for spin-up and spin-down in-gap
states are shifted oppositely by the ISOC, and the sign of the
relative shifts of them depends on which sublattices the site is
belonging to. Based on a scenario of local breaking of the spin and
sublattice degeneracies due to the interaction of the ISOC derived
effective fluxes with the local magnetic flux inside the vortex
core, we give an explanation to the unusual phenomena regarding the
polarization of the induced magnetic orders and the energy shifts of
the finite-energy in-gap peaks. The calculated results may not only
serve as experimental signatures for identifying the ISOC proposed
in the Ising superconductors, but also put forward effective
thinking-ways in manipulation of electron spins in motion through
the orbital effect in the SC vortex states.

The remainder of the paper is organized as follows. In Sec. II, we
introduce the model Hamiltonian and carry out analytical
calculations. In Sec. III, we present numerical calculations and
discuss the results. In Sec. IV, we make a conclusion.

\section{THEORY AND METHOD}
\begin{figure}[htb]
\begin{center}
\vspace{-.2cm}
\includegraphics[width=240pt,height=200pt]{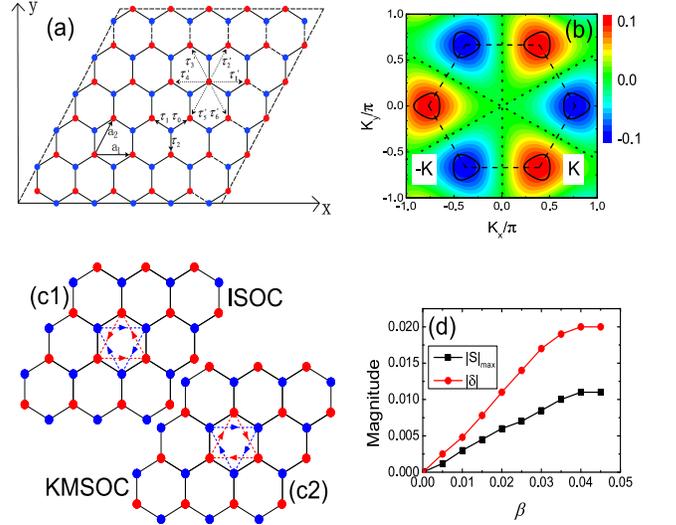}
\caption{(a) Honeycomb lattice structure of the Ising
superconductor, made out of two sublattices $A$ (blue dots) and $B$
(red dots). $\tau_{0}$, $\tau_{1}$ and $\tau_{2}$ are the
nearest-neighbor vectors, and $\tau'_{1}$-$\tau'_{6}$ the
next-nearest-neighbor vectors. (b) The Brillouin zone (dashed line)
and the six spin degenerate Fermi pockets (solid lines) of the Ising
superconductor. The red and blue colors indicate the opposite sign
of the effective Zeeman fields between adjacent Fermi pockets
located at $\mathbf{K}$ and $-\mathbf{K}$. The positive phase
hopping directions for spin-up electrons depicted by $H_{ISOC}$ in
Eq. (6) (c1), and by $H_{KM}$ in Eq. (15) (c2), respectively. The
arrows in both figures indicate the positive phase hopping
directions. (d) The ISOC dependencies of the maximum of the absolute
value for the induced magnetic order $|S|_{max}$ and the magnitude
of the relative energy shifts $|\delta|$ between the spin-up and
spin-down in-gap state peaks on the vortex core center [reference to
text and Fig.~\ref{fig4}(b)].}\label{fig1}
\end{center}
\end{figure}

The effective electron hoppings between the NN sites $i$ and
$i+\tau_{j}$ on a honeycomb lattice can be described by the
following tight-binding Hamiltonian,
\begin{eqnarray}
H_{0}&=&-\sum_{i,\tau_{j},\sigma}(t_{i,i+\tau_{j}}a^{\dag}_{i,\sigma}b_{i+\tau_{j},\sigma}+h.c.)
-\mu(\sum_{i\in A,\sigma}a^{\dag}_{i,\sigma}a_{i,\sigma}\nonumber\\
&& +\sum_{i\in B,\sigma}b^{\dag}_{i,\sigma}b_{i,\sigma}),
\end{eqnarray}
where $t_{i,i+\tau_{j}}$ is the hopping integral between the NN
sites. $\mathbf{\tau}_{j}$ denotes the three NN vectors with
$\mathbf{\tau}_{0}=a(\frac{\sqrt{3}}{2},\frac{1}{2})$,
$\mathbf{\tau}_{1}=a(-\frac{\sqrt{3}}{2},\frac{1}{2})$ and
$\mathbf{\tau}_{2}=a(0,-1)$ as defined in Fig.~\ref{fig1}(a) with
$a$ being the lattice constant. $a^{\dag}_{i,\sigma}
(b^{\dag}_{i,\sigma})$ is the electron creation operator in
sublattice $A$ $(B)$ if $i\in$ sublattice $A$ $(B)$, and $\mu$ the
chemical potential. For the free hopping case with
$t_{i,i+\tau_{j}}=t$, the Hamiltonian $H_{0}$ can be written in the
momentum space,
\begin{eqnarray}
H_{0}(k)&=&\sum_{k,\sigma}[\xi_{k}a^{\dag}_{k,\sigma}b_{k,\sigma}
+\xi^{\ast}_{k}b^{\dag}_{k,\sigma}a_{k,\sigma}-\mu
(a^{\dag}_{k,\sigma}a_{k,\sigma} \nonumber\\
&&+b^{\dag}_{k,\sigma}b_{k,\sigma}) ],
\end{eqnarray}
where
\begin{eqnarray}
\xi_{k}=-t\sum_{j=0}^{2}e^{i\mathbf{k}\cdot\mathbf{\tau}_{j}}.
\end{eqnarray}
One can readily find the energy bands for this Hamiltonian
as~\cite{neto1},
\begin{eqnarray}
\varepsilon^{\pm}_{k}&=&\pm
t[3+2\cos(\sqrt{3}k_{x})+4\cos(\sqrt{3}k_{x}/2)\cos(3k_{y}/2)]^{\frac{1}{2}} \nonumber\\
&&-\mu.
\end{eqnarray}
with $+$ ($-$) indexing the conduction (valence) band. We focus on
systems which have been doped such that the chemical potential $\mu$
lies in the upper conduction bands, and produce six spin degenerate
pockets at the corners of the hexagonal Brillouin zone when
$\varepsilon^{+}_{k}=0$, as shown in Fig.~\ref{fig1}(b).

\vspace*{.2cm}

The ISOC acts as strong effective Zeeman fields, which polarize
electron spins oppositely to the out-of-plane direction at opposite
valleys, that is, at the $\mathbf{K}$ and $-\mathbf{K}$ points in
Fig.~\ref{fig1}(b). If we choose the out-of-plane direction as the
$z$-axis, the ISOC term has the form~\cite{Frigeri1}
\begin{eqnarray}
H_{ISOC}(k)=\beta\sum_{k,\sigma,\sigma'}\mathbf{g}_{k}\cdot\mathbf{\hat{\sigma}}_{\sigma\sigma'}
(a^{\dag}_{k,\sigma}a_{k,\sigma'}
+b^{\dag}_{k,\sigma}b_{k,\sigma'}),
\end{eqnarray}
where $\beta$ is the ISOC strength, and $\hat{\sigma}$ denotes the
Pauli matrices acting in the spin space. The ISOC requires that the
form factor $\mathbf{g}_{k}$ alternates its sign between adjacent
Fermi pockets located at $\mathbf{K}$ and $-\mathbf{K}$ [see
Fig.~\ref{fig1}(b)], which should be the form
$\mathbf{g}_{k}=\hat{z}F_{k}$ with
$F_{k}=2\sin(\sqrt{3}k_{x})-4\cos(3k_{y}/2)\sin(\sqrt{3}k_{x}/2)=-F_{-k}$
satisfying the time-reversal symmetry. In this way, the spins are
bound to the orbitals in the momentum space and accordingly exhibit
various valley dependent behaviors such as valley spintronics in
these materials~\cite{Radisavljevic,Zhang1,Wang1,Bao1,Lee1}. By
making the Fourier transformation of $F_{k}$, the ISOC term in real
space can be reached as~\cite{Jiang1},
\begin{eqnarray}
H_{ISOC}=i\beta\sum_{i,\mathbf{\tau}'_{j},\sigma,\sigma'}\hat{\sigma}^{z}_{\sigma\sigma'}
(-1)^{j} (a^{\dag}_{i,\sigma}a_{i+\mathbf{\tau}'_{j},\sigma'}
+b^{\dag}_{i,\sigma}b_{i+\mathbf{\tau}'_{j},\sigma'}),
\end{eqnarray}
where the vectors $\mathbf{\tau}'_{j}$ connecting the six
next-nearest-neighbor (NNN) sites are located at
$\mathbf{\tau}'_{1}=-\mathbf{\tau}'_{4}=\sqrt{3}a(1,0)$,
$\mathbf{\tau}'_{2}=-\mathbf{\tau}'_{5}=\sqrt{3}a(\frac{1}{2},\frac{\sqrt{3}}{2})$
and
$\mathbf{\tau}'_{3}=-\mathbf{\tau}'_{6}=\sqrt{3}a(-\frac{1}{2},\frac{\sqrt{3}}{2})$,
as indicated by the dashed arrows in Fig.~\ref{fig1}(a). We will see
later that the ISOC in real space depicted by Eq. (6) plays the role
of the coupling between spins and the effectively periodic fluxes
with the quantization axis along the out-of-plane direction. Then,
the Hamiltonian including both the free hoppings and the ISOC term
is reached, in real space as,
\begin{eqnarray}
H_{TMD}=H_{0}+H_{ISOC}.
\end{eqnarray}

The SC pairing is assumed to be derived from the effective
attraction between electrons,
\begin{eqnarray}
H_{P}&=&\frac{V_{0}}{2}\sum_{i,\sigma}n_{i,\sigma}n_{i,\bar{\sigma}}.
\end{eqnarray}
Here, we consider the on-site interactions with $V_{0}$ denoting the
effective interaction potential~\cite{btzhou1,Sharma1}. By making
the mean-field decoupling, $H_{P}$ can be rewritten in terms of the
SC pairings as,
\begin{eqnarray}
H_{P}&=&\sum_{i\in
A}(\Delta_{A}a^{\dag}_{i,\uparrow}a^{\dag}_{i,\downarrow}+h.c.)
+\sum_{i\in
B}(\Delta_{B}b^{\dag}_{i,\uparrow}b^{\dag}_{i,\downarrow} \nonumber\\
&&+h.c.),
\end{eqnarray}
where $\Delta_{A}=-V_{0}\langle
a_{i,\uparrow}a_{i,\downarrow}\rangle$ ($\Delta_{B}=-V_{0}\langle
b_{i,\uparrow}b_{i,\downarrow}\rangle$) defines the on-site
spin-singlet $s$-wave SC pairing.

Then the total Hamiltonian is arrived as follows,
\begin{eqnarray}
H=H_{TMD}+H_{pair}.
\end{eqnarray}
Based on the Bogoliubov transformation, the diagonalization of the
Hamiltonian $H$ can be achieved by solving the following discrete
BdG equations,
\begin{eqnarray}
\sum_{j}\left(
\begin{array}{cccc}
-\mu\delta_{ij} & H_{ij,\uparrow\uparrow} &
\Delta_{A}\delta_{ij} & 0 \\
H^{\ast}_{ij,\uparrow\uparrow} & -\mu\delta_{ij} & 0 & \Delta_{B}\delta_{ij} \\
\Delta^{\ast}_{A}\delta_{ij} & 0 &
\mu\delta_{ij} & -H^{\ast}_{ij,\downarrow\downarrow} \\
0 & \Delta^{\ast}_{B}\delta_{ij} & -H_{ij,\downarrow\downarrow} &
\mu\delta_{ij}
\end{array}
\right)\left(
\begin{array}{cccc}
u_{A,n,j,\uparrow} \\
u_{B,n,j,\uparrow} \\
v_{A,n,j,\downarrow} \\
v_{B,n,j,\downarrow}
\end{array}
\right)= \nonumber\\ E_{n}\left(
\begin{array}{cccc}
u_{A,n,i,\uparrow} \\
u_{B,n,i,\uparrow} \\
v_{A,n,i,\downarrow} \\
v_{B,n,i,\downarrow}
\end{array}
\right),
\end{eqnarray}
where,
\begin{eqnarray}
H_{ij,\uparrow\uparrow}=-t_{ij}\delta_{i+\mathbf{\tau}_{j},j}+i\beta\sigma^{z}_{\uparrow\uparrow}(-1)^{j}\delta_{i+\mathbf{\tau}'_{j},j},  \nonumber\\
H_{ij,\downarrow\downarrow}=-t_{ij}\delta_{i+\mathbf{\tau}_{j},j}+i\beta\sigma^{z}_{\downarrow\downarrow}(-1)^{j}\delta_{i+\mathbf{\tau}'_{j},j},
\end{eqnarray}
with $u_{A,n,j,\uparrow}$ ($u_{B,n,j,\uparrow}$) and
$v_{A,n,j,\downarrow}$ ($v_{B,n,j,\downarrow}$) being the Bogoliubov
quasiparticle amplitudes on the $j$-th site with corresponding
eigenvalues $E_{n}$. The SC pairing amplitudes satisfy the following
self-consistent conditions,
\begin{eqnarray}
\Delta_{A}=-\frac{V_{0}}{2}\sum_{n}u_{A,n,i,\uparrow}
v^{\ast}_{A,n,i,\downarrow}\tanh(\frac{E_{n}}{2k_{B}T}), \nonumber\\
\Delta_{B}=-\frac{V_{0}}{2}\sum_{n}u_{B,n,i,\uparrow}
v^{\ast}_{B,n,i,\downarrow}\tanh(\frac{E_{n}}{2k_{B}T}).
\end{eqnarray}

The spin dependent electron density $n_{A(B),i,\sigma}$ and the
local magnetic orders $S_{A(B),i,z}$ are determined respectively by,
\begin{eqnarray}
n_{A(B),i,\uparrow}=\sum_{n}|u_{A(B),n,i,\uparrow}|^{2}
f(E_{n}), \nonumber\\
n_{A(B),i,\downarrow}=\sum_{n}|v_{A(B),n,i,\downarrow}|^{2}
f(E_{n}), \nonumber\\
S_{A(B),i,z}=\frac{1}{2}[n_{A(B),i,\uparrow}-n_{A(B),i,\downarrow}].
\end{eqnarray}

\section{results and discussion}

In numerical calculations, we choose the zero field hopping integral
$t=200$meV as the energy unit, and fix temperature $T=1\times
10^{-5}$, unless otherwise specified. The filling factor
$n=\sum_{i,\sigma}n_{i,\sigma}/N=1.08$ ($N$ denotes the number of
total lattice sites) such that the chemical potential $\mu$ lies in
the upper conduction band and gives rise to the Fermi surfaces in
Fig.~\ref{fig1}(b). In the presence of a perpendicular magnetic
field, the orbital effect dominates over the the Zeeman splitting,
so we neglect the Zeeman term of the external magnetic field in the
following calculations. In this case, the hopping terms are
described by the Peierls substitution. For the NN hopping between
sites $i$ and $i+\mathbf{\tau}_{j}$, one has
$t_{i,i+\mathbf{\tau}_{j}}=te^{i\varphi_{i,i+\mathbf{\tau}_{j}}}$,
and for the NNN hopping between $i$ and $i+\mathbf{\tau}'_{j}$ one
should have $\beta\rightarrow \beta
e^{i\varphi_{i,i+\mathbf{\tau}'_{j}}}$, where
$\varphi_{i,i+\mathbf{\tau}_{j}(\mathbf{\tau}'_{j})}=
\frac{\pi}{\Phi_{0}}\int^{r_{i}}_{r_{i+\mathbf{\tau}_{j}(\mathbf{\tau}'_{j})}}\mathbf{A}(\mathbf{r})\cdot
d\mathbf{r}$ with $\Phi_{0}=\frac{hc}{2e}$ being the SC flux quanta.
We consider a system with a parallelogram vortex unit cell as shown
in Fig.~\ref{fig1}(a), where two vortices are accommodated. The
vortex unit cell with size of
$24\mathbf{a_{1}}\times48\mathbf{a_{2}}$ is adopted in the
calculations, unless otherwise stated. The vector potential
$\mathbf{A}(\mathbf{r})=(0,Bx,0)$ is chosen in the Landau gauge to
give rise to the magnetic field $\mathbf{B}$ along the
$z$-direction.

In this study, we have no ambition to explore the SC mechanism
underlying the Ising superconductors. Instead, we assume a
phenomenological pairing potential $V_{0}$ to give rise to the SC
pairing. Within the BCS theory, the coherence length is given by
$\xi_{0}=\hbar v_{F}/\pi\Delta$, where $v_{F}$ is the Fermi
velocity, linking the coherence length to the inverse size of the SC
gap $\Delta$. The coherence length of NbSe$_{2}$ is about $10$nm as
obtained from $H_{c2}(T)$ measurement~\cite{Fente1,Kogan1}. The
estimated vortex core size is of $\xi_{V}\sim 30$nm~\cite{Fente1}. A
system contains two such vortex cores would be larger than the size
of $60$nm$\times 120$nm, which roughly amounts to a parallelogram
sample with the size larger than $200\mathbf{a}_{1}\times
400\mathbf{a}_{2}$. Such a large size is far beyond the
computational capability. However, it is still capable of mimicking
the vortex physics on a relative small size of sample by
artificially enlarging the SC gap $\Delta$. In the self-consistent
calculations, the length scale of the sample with size
$24\mathbf{a}_{1}\times 48\mathbf{a}_{2}$ is about one order smaller
than the actual size. Thus, we need to choose a large $V_{0}=1.6$ in
the self-consistent calculations to give rise to a bulk value of
$\Delta\approx 0.09\sim 18$meV, a value about one order larger than
the actual measurements~\cite{Fente1}, so as to meet the
requirement.

\vspace*{.2cm}
\begin{figure}[htb]
\begin{center}
\vspace{-.2cm}
\includegraphics[width=250pt,height=170pt]{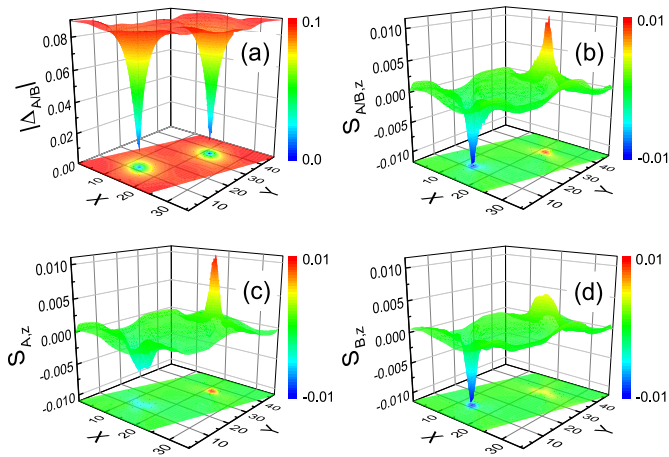}
\caption{The spatial distributions of the SC and magnetic order
parameters in the vortex states for $\beta=0.04$ are shown in (a)
and (b), respectively. The spatial distributions of the magnetic
order in the vortex states for $\beta=0.04$ on sublattice $A$ (c),
and on sublattice $B$ (d), respectively.}\label{fig2}
\end{center}
\end{figure}

\subsection{The induced antiphase magnetic orders inside vortex cores}
Under a perpendicular magnetic field, the vanishment of the
screening current density at the vortex center drives the system
into the vortex states with the suppression of the SC order
parameter around the vortex core. In the absence of the ISOC
interaction, we find that except for the suppression of the SC order
around the vortex core region there is no other order to be induced.
On the other hand, when the ISOC is present, a ferromagnetic order
can emerge inside the vortex core region with its maximum appearing
at the vortex core center. The maximum of the absolute value for the
magnetic order $|S|_{max}$ exhibits roughly linear increasing trend
with $\beta$ in a wide range of ISOC, and finally reaches a
saturated value at large ISOC, as displayed in Fig.~\ref{fig1}(d).
Typical results on the vortex structure with $\beta=0.04$ are shown
in Figs.~\ref{fig2}(a) and~\ref{fig2}(b) for the spatial
distributions of SC and magnetic orders, respectively. As shown in
Fig.~\ref{fig2}(a), each vortex unit cell accommodates two SC
vortices each carrying a flux quantum $\Phi_{0}$. The SC order
parameter $|\Delta_{A/B}|$ vanishes at the vortex core center where
the maximum of the induced magnetic order appears. It is interesting
to note that the magnetic order parameters have opposite polar
directions around two NN vortices along the long side of the
parallelogram vortex unit cell, as shown in Fig.~\ref{fig2}(b). The
most unusual aspect of the spatial distribution of the magnetic
order parameters $S_{A(B),i,z}$ appears when we replot in
Figs.~\ref{fig2}(c) and~\ref{fig2}(d) the magnetic orders separately
on the sublattices $A$ and $B$. Specifically, the positive magnetic
order alone $z$-axis inside one vortex comes dominantly from the $A$
sublattice while the negative one inside another vortex comes
dominantly from the $B$ sublattice.

In order to understand the origin as well as the unusual
distributions of the induced magnetic order, we should note the fact
that there is no magnetic order induced when ISOC is zero. In real
space, the ISOC depicted by Eq. (6) plays the role of coupling
between spins and the effectively periodic fluxes with the
quantization axis along the out-of-plane direction. Following the
ISOC term in Eq. (6), we display the positive phase [noting that
$i=e^{i\pi/2}$] hopping directions of ISOC in Fig.~\ref{fig1}(c1) by
arrows on NNN bonds for spin-up electrons at sublattices $A$ and
$B$, from which the effective spin fluxes are generated. If the
positive phase hoppings on NNN bonds for spin-up electrons on
sublattice $A$ generate spin flux pointing to $z$-direction, then
they generate spin flux pointing to $-z$-direction on sublattice
$B$, and contrary is true for spin-down electrons. That is, the NNN
hoppings have opposite chirality, for sublattices $A$ and $B$. Since
the spins, sublattices and the effectively periodic fluxes are bound
together in real space, local breaking of the spin and sublattice
degeneracies may be expected if the effective fluxes for sublattices
$A$ and $B$ are contrastively altered by an out-of-plane magnetic
field, and thus the spin orders in real space may also be expected.
Nevertheless, we can not expect the appearance of magnetic order in
the normal state under an out-of-plane magnetic field. This is due
to the fact that the energy scale of the hopping integral $t$
overwhelms the ISOC strength $\beta$, interchanging the electrons
between sites of sublattices $A$ and $B$ leading to the suppression
of the local orders. However, the situation is totally different in
the vortex state, where the localized electrons in the vortex core,
which come from the breaking of the Cooper pairs, contribute to the
magnetic order. If one vortex core resides on the $A$ sublattice
site, the blue site shown in Fig.~\ref{fig1}(a), the positive phase
hoppings on NNN bonds bound to spin-up electrons on sublattice $A$
generate effective spin flux pointing to $z$-direction [noting the
negative charge of the electrons], which is in the same direction as
the magnetic field. On the contrary, the spin-down electrons on
sublattice $A$ generate effective spin flux in the opposite
direction of the magnetic field. Thus, the spin degeneracy breaks
locally to two branches with a lower energy for the spin-up
electrons, leading to the positive magnetic order around one vortex
as shown in Fig.~\ref{fig2}(c) on sublattice $A$. In principle, the
pairing breaking from the spin-singlet SC pairings due to the
orbital effect of the magnetic field results in equal numbers of
spin-up and spin-down electrons, so the total spins should be zero
as a global. The excess of spin-down electrons accumulate into the
NN vortex to give rise to the negative magnetic order shown in
Fig.~\ref{fig2}(d) on sublattice $B$, whereby it saves the energy as
the effective spin flux generated by spin-down electrons being in
compliance with the direction of the magnetic field.

\begin{figure}[htb]
\begin{center}
\vspace{0.8cm}
\includegraphics[width=250pt,height=170pt]{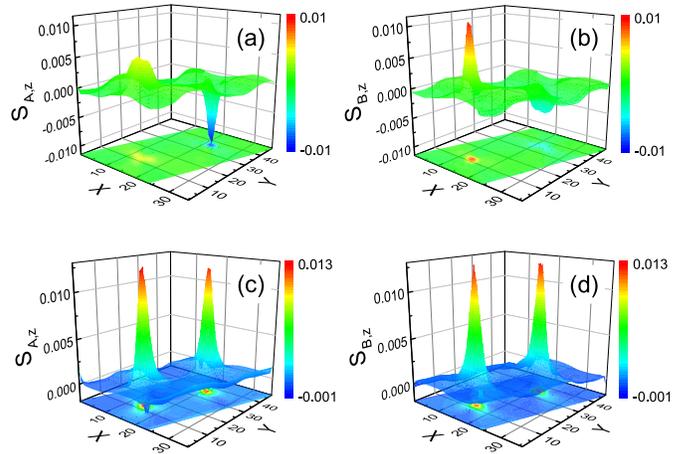}
\caption{The spatial distributions of the induced magnetic orders in
the vortex states with the magnetic field along the $-z$ direction
for $\beta=0.04$ on sublattice $A$ (a), and on sublattice $B$ (b),
respectively. (c) and (d) show the calculated results fot the
spatial distributions of the induced magnetic orders by replacing
$H_{ISOC}$ with $H_{KM}$ [see text for details].}\label{fig3}
\end{center}
\end{figure}

Two situations could lend support to the above scenario. Firstly, we
consider the case with a reversal of the direction of the magnetic
field, i.e., a magnetic field in the $-z$-direction. From the above
argument, the polarizations of the induced magnetic orders should be
reversed if the magnetic field reverses its direction. It is exactly
the case as evidenced in Figs.~\ref{fig3}(a) and~\ref{fig3}(b),
where the results are obtained with an out-of-plane magnetic field
in the $-z$-direction while keep other parameters the same as that
in Fig.~\ref{fig2}. Secondly, we should make a comparison with the
spin-orbital coupling (SOC) term in Kane-Mele model~\cite{Kane1},
which has the form
\begin{eqnarray}
H_{KM}=i\beta\sum_{i,\mathbf{\tau}'_{j},\sigma,\sigma'}\hat{\sigma}^{z}_{\sigma\sigma'}
(-1)^{j}(a^{\dag}_{i,\sigma}a_{i+\mathbf{\tau}'_{j},\sigma'}
-b^{\dag}_{i,\sigma}b_{i+\mathbf{\tau}'_{j},\sigma'}).
\end{eqnarray}
Both $H_{ISOC}$ and $H_{KM}$ preserve time-reversal symmetry, so the
spins remain degenerate in both cases. The only difference lies that
$H_{ISOC}$ preserves the sublattice symmetry but $H_{KM}$ breaks it.
As a result, the NNN hopping phases carried by the same spins in
$H_{KM}$ would have same chirality for sublattices $A$ and $B$, as
denoted by arrows in Fig.~\ref{fig2}(c2). According to the above
scenario, we deduce that the induced magnetic orders should be in
the same direction for the two adjacent vortices. This is also
verified in Figs.~\ref{fig3}(c) and~\ref{fig3}(d), where the results
for the spatial distribution of the induced magnetic orders are
calculated by replacing $H_{ISOC}$ with $H_{KM}$ while other
parameters remain unchanged.

\vspace*{-.0cm}
\vspace*{-.0cm}

\subsection{The splitting and shift of the finite-energy peaks for
the spin-resolved LDOS} 
\begin{figure}[htb]
\begin{center}
\vspace{0.8cm}
\includegraphics[width=250pt,height=210pt]{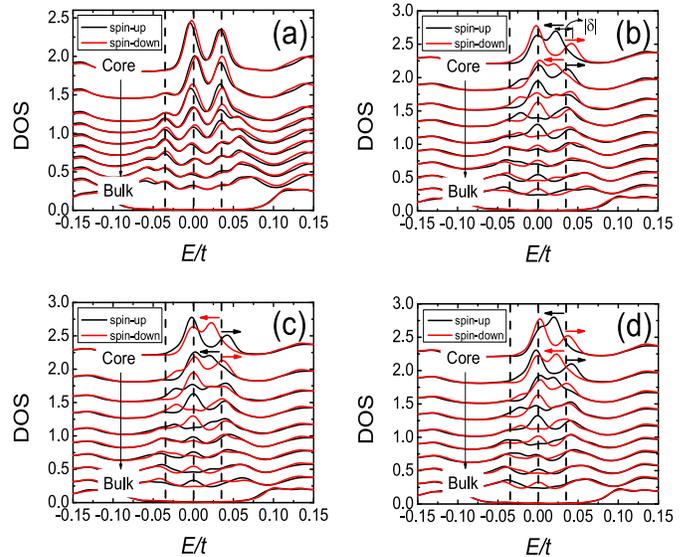}
\caption{The energy dependence of the spin-resolved LDOS on a series
of sites for $\beta=0.0$ (a), and for $\beta=0.04$ (b), (c) and (d).
(a), (b) and (d) are the results for a magnetic field along the
$z$-direction, while (c) the results for a magnetic field along the
$-z$-direction. (b) and (c) show the LDOS inside the same vortex
core, and (d) the LDOS inside another vortex core. In each panel
from top to bottom, the curves stand for the LDOS at sites along the
zigzag direction moving away from the core center. The curves are
vertically shifted for clarity. The three dashed vertical lines in
each panel denote the three low energy peak positions for $\beta=0$.
The arrows in (b), (c) and (d) indicate the peak position shift with
respect to that of $\beta=0$. The magnitude of the relative energy
shifts $|\delta|$ is shown in (b).}\label{fig4}
\end{center}
\end{figure}

Next, we examine the energy dependence of
the LDOS in the vortex states on the honeycomb lattice. The LDOS is
defined as
$N(\mathbf{R}_{i},E)=N_{\uparrow}(\mathbf{R}_{i},E)+N_{\downarrow}(\mathbf{R}_{i},E)$
with
$N_{\uparrow}(\mathbf{R}_{i},E)=-\sum_{n}|u_{A(B),n,i,\uparrow}|^{2}f'(E_{n}-E)$
and
$N_{\downarrow}(\mathbf{R}_{i},E)=-|v_{A(B),n,i,\downarrow}|^{2}f'(E_{n}+E)$
being the spin-resolved LDOS for spin-up and spin-down states,
respectively. In order to reduce the finite size effect, the
calculations of the LDOS are carried out on a periodic lattice which
consists of $16\times 8$ parallelogram vortex unit cells, with each
vortex unit cell being the size of
$24\mathbf{a_{1}}\times48\mathbf{a_{2}}$. In Fig.~\ref{fig4}, we
plot a series of the spin-resolved LDOS as a function of energy at
sites along the zigzag direction moving away from the vortex center
for $\beta=0.0$ and $\beta=0.04$, respectively. For comparison, we
have also displayed the LDOS at the midpoint between the two NN
vortices, which resembles the U-shaped full gap feature for the bulk
system. In the absence of ISOC, the states of spin-up and spin-down
are nearly equal occupation and empty in the vortex core, as shown
in Fig.~\ref{fig4}(a), in accordance with the empty cores without
the induced magnetic orders. Besides the almost identical LDOS line
shapes for the spin-up and spin-down states, the LDOS shown in
Fig.~\ref{fig4}(a) exhibits another two prominent features within
the SC gap edges. On one hand, the LDOS shows the pronounced
discrete energy peaks inside the core region with one located near
the zero energy and others located at finite energies, as indicated
by the dashed vertical lines in the figure. Here, the asymmetric
line shape of the LDOS with respect to zero energy reflects the lack
of particle-hole symmetry as the chemical potential $\mu$ deviates
from zero for the filling factor $n$ being greater than the half
filling ($n>1$). Due to the particle-hole asymmetry, the
finite-energy bound states at the core site only appear on the $E>0$
side~\cite{Haya1} (There are also weak peaks at finite energies on
the $E<0$ side when moving away from the core center.). The
existence of the zero-energy vortex core sates in the Dirac fermion
system have been predicted analytically by Jackiw and Rossi in terms
of the zero-energy solutions of relativistic field
theory~\cite{Jack1}. Although these zero-energy solutions were
subsequently demonstrated that the existence of these zero-energy
solutions is connected to an index theorem~\cite{Wein1} and the zero
modes were shown to exist in the Dirac continuum theory of the
honeycomb lattice at half filling~\cite{Ghaemi1}, the zero-energy
levels split when adopting a honeycomb lattice model description by
setting the size of the vortex core to be zero~\cite{Doro1}. It is
also found that the energy splitting decreases with the vortex size
and leads to the near-zero-energy states in the circumstance of
finite core size~\cite{Doro1}. While the notion of the zero-energy
vortex core states presents an important subject of study being
worthy of further research, we identify the near-zero-energy vortex
core states here in a self-consistent manner by employing the
honeycomb lattice model, where the band structure has the Dirac-type
dispersion near the half filling. On the other hand, though the
peaks' intensities are suppressed as the site departing from the
core center, the energy levels of these peaks are almost independent
of positions. It is worth while to note that a dispersionless
zero-energy conductance peak has been recently observed inside the
SC vortex core by Chen's group~\cite{Liang1} in the kagome
superconductor CsV$_{3}$Sb$_{5}$, which shares the lattice structure
with component of hexagonal honeycomb and the electronic structure
with Dirac points in a manner similar to those in honeycomb
lattices. How the calculated results with near-zero-energy peaks in
the present study relate to the experimental observations, and
whether these near-zero-energy vortex core states have a common
underlying symmetrical cause, constituting another fascinating
questions deserving further studies.

\vspace*{-.0cm}
\vspace*{-.0cm}

In the presence of the ISOC, the local breaking of the spin and
sublattice degeneracies in the vortex states is also reflected in
the energy dependence of the LDOS.
Figs.~\ref{fig4}(b),~\ref{fig4}(c) and~\ref{fig4}(d) present the
typical results of the spin-resolved LDOS for $\beta=0.04$. As can
be seen from Fig.~\ref{fig4}(b), while the energy level of the
near-zero-energy peaks remain virtually unchanged for both spins,
the energy levels of the finite-energy peaks are shifted differently
by the ISOC for different spins and at different sublattice sites,
as compared with the case of $\beta=0$. Specifically, for the LDOS
on the same site within the core region, the finite-energy peaks for
the spin-up and spin-down bound states shift oppositely, as
indicated by the arrows in the figures, depicting a picture of local
breaking of the spin degeneracy. At the same time, for the bound
states with the same spin, the finite-energy peaks on the sites
belonging to different sublattices also have the opposite shifts,
indicating the local breaking of the sublattice degeneracy. Since
there are induced magnetic orders in the vortex cores as well as the
similar ISOC dependencies of the magnitudes of the magnetic orders
$|S|_{max}$ and the relative energy shifts $|\delta|$ as shown in
Fig.~\ref{fig1}(d), it is natural to suspect whether the spin
splitting for the LDOS is derived from the Zeeman effect of the
induced local magnetic order interacting with the
electrons~\cite{Zhu2}, or from the above scenario where the spin
degree of freedom is manipulated by the orbital effect of magnetic
field via the ISOC. Several aspects render the Zeeman effect
mechanism impossible. As has been shown in Fig.~\ref{fig2}(b), the
magnetic orders polarize oppositely inside two NN vortices. If the
Zeeman effect mechanism runs, the energy level shifts of the peaks
should behave the opposite way on the sites located respectively at
the two NN vortices. Nevertheless, as displayed in
Figs.~\ref{fig4}(b) and~\ref{fig4}(d), the consistency of the peaks'
shifts on the sites located at different vortices while belonging to
the same sublattice rules out the Zeeman effect mechanism. The
second thing we notice about the energy level shifts of the peaks is
that they occur only for the ones with finite energy, while the
near-zero-energy peaks almost stay the same, being at odds with the
Zeeman effect mechanism. Finally, if we reverse the direction of the
out-of-plane magnetic field, as shown in Fig.~\ref{fig4}(c), the
peaks' shifts behave exactly the opposite way as compared with that
in Fig.~\ref{fig4}(b). It is thus confirmed that the local spin
splitting and the local breaking of sublattice degeneracy are
conformed with the above scenario where the spin degree of freedom
is manipulated by the orbital effect of magnetic field via the ISOC
in the SC vortex states.

\vspace*{.2cm}
\begin{figure}[htb]
\begin{center}
\vspace{-.2cm}
\includegraphics[width=240pt,height=90pt]{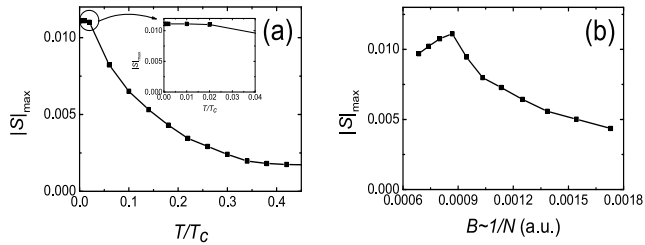}
\caption{Temperature (a), and magnetic field (b) evolutions of the
maximum of the absolute value for the induced magnetic orders at vortex cores with $\beta=0.04$.}\label{fig5}
\end{center}
\end{figure}

\subsection{The effects of temperature and magnetic field strength on the induced orders}
Due to the 2D nature of the Ising superconductors, the thermal
effect on the induced magnetic orders constitutes an inevitable
issue from both theoretical perspective and experimental
realization. Although the system under study is 2D, the induced
magnetic orders are formed under the combined actions of the
magnetic field, the ISOC and the SC order, so they are not
spontaneous ones. Meanwhile, the induced magnetic orders are
localized inside the vortex core regions, and thus they are local
ones. As a result, one may expect a different manner of the thermal
effect on the induced magnetic orders as compared with the
Mermin-Wagner theorem~\cite{Mermin1}. To see the thermal effect on
the induced magnetic orders, we calculate the temperature dependence
of the magnitude of the magnetic orders. Fig.~\ref{fig5}(a) shows
the temperature dependence of the maximum of the absolute value for
the induced magnetic order at the vortex core, where $T$ is rescaled
by $T_{c}\approx 0.05$. As can be seen from the figure, the
magnitude of the magnetic orders remains approximately constant at
low temperature $T\leq 0.02T_{c}$ [see inset of Fig.~\ref{fig5}(a)]
as a result of the small thermal excitations and the almost
unchanged vortex core size at this temperature
regime~\cite{Miller1}. After then it shows a steady decreasing trend
with increasing temperature, and finally reaches a tiny value at
$T\sim 0.5T_{c}$. The decreasing trend is mainly ascribed to the
enlarging vortex core size with temperature~\cite{Miller1}, the
so-called Kramer-Pesch Effect~\cite{Kramer1}. The enlarged vortex
core would involve more different sublattice sites into the vortex
core center, resulting in the reduction of the induced magnetic
orders. Though the magnetic orders reduce their magnitude upon the
increasing of the temperature, they sustain to a finite temperature.
Therefore, one may expect to observe the induced magnetic orders
under temperatures well below the SC critical temperature.

Another important factor to be considered in observing the induced
magnetic orders is how the strength of the external magnetic field
affects the induced magnetic orders. Since one vortex unit cell
accommodates two vortices in the calculations, we have
$B=2\Phi_{0}/A\sim 1/N$ with $A$ and $N$ being the area and the site
number of the vortex unit cell. Fig.~\ref{fig5}(b) displays the
variation of the maximum of the absolute value for the induced
magnetic orders with respect to different strengths of the magnetic
field, which are realized in the calculations by varying the size of
the parallelogram vortex unit cell. In the weak to moderate magnetic
field region, there is little interference between the vortex cores
due to the large inter-vortex spacing $d$. The increase of the
magnetic field leads to more broken Cooper pairs inside the vortex
cores to contribute to the formation of the magnetic orders, so the
magnitude of the induced magnetic orders increases with the magnetic
field strength, as evidenced in Fig.~\ref{fig5}(b). However, as the
magnetic field increasing further, the adjacent vortex cores with
opposite polarizations of the induced magnetic order would get close
enough (with a length scale being less than two times of the
penetration depth $\lambda$) to interfere with one another, leading
to the reduction of the magnitude of the magnetic order. This
suggests the induced magnetic orders will be altered in an Abrikosov
vortex lattice~\cite{Abrik1}. On one hand, the formation of the
Bloch wave~\cite{Franz1} or the interactions among
vortices~\cite{Blatter1} in the vortex lattice will suppress the
induced magnetic orders. On the other hand, since there are many
vortices in the sample instead of just two, the polarization of the
induced orders is not necessarily opposite for two adjacent vortex
cores. Nevertheless, the result also means the induced magnetic
orders would survive in the vortex lattice under a weak to moderate
magnetic field as long as $d\gg\lambda$, i.e., the inter-vortex
spacing is much larger than the penetration depth.

\section {remarks and conclusion}
The local magnetic orders induced in the SC vortex states have been
extensively investigated on the cuprates
superconductors~\cite{Ogata1,Zhu2,Zhu3,Chen1,Taki1,Tsuch1}, where
the emergence of the magnetic orders inside the core region was
generally believed to be originated from the electrons'
correlations. These correlations usually come from the Coulomb
interactions between electrons and that the induced magnetic orders
have nothing to do with the chirality of the electrons. However, the
induced local magnetic orders inside the SC vortex core by the ISOC
has a direct bearing on what the electrons' chirality is. As has
been demonstrated that the amplitudes of the induced magnetic orders
and the unusual energy shifts of the in-gap state peaks present here
are related to the ISOC strength $\beta$, while their directions are
determined by the direction of the magnetic field. The amplitude and
the different polar direction of the induced local magnetic orders
could be measured by the muon spin rotation ($\mu$SR) spectroscopy
and the nuclear magnetic resonance experiments, and the energy
shifts of the in-gap state peaks on different sublattice for
different spins could be observed in the spin-polarized scanning
tunneling microscopy experiments. Both of these observations may be
served as signatures to characterize the ISOC proposed for the Ising
superconductors. In the meantime, since the induced magnetic orders
are derived from the ISOC, the breaking of the spin degeneracy and
the energy shifts of the in-gap state peaks are selectively occurred
for the electrons which possess finite momentum with respect to the
vortex center. The scenario proposed here may also provide a
possibility in manipulation of electron spins in motion via the
orbital effect in the SC vortex states.

In conclusion, we have numerically investigated the vortex states of
the Ising superconductors, with the emphasis on the local breaking
of the spin and sublattice degeneracies as a result of the
interaction between the ISOC derived effective fluxes and the local
magnetic flux inside the vortex core. In the absence of the ISOC,
there was no magnetic order induced inside the vortex core, and the
almost identical line shapes of the LDOS for the spin-up and
spin-down in-gap states were shown up inside the core region,
forming a series of discrete energy peaks within the gap edges. The
inclusion of the ISOC induced the ferromagnetic orders inside the
vortex core region, where the magnetic orders polarized oppositely
for the two NN vortices with one of the two polarizations coming
dominantly from one specie of the two sublattices. Accordingly, the
finite-energy peaks of the LDOS on the same site for spin-up and
spin-down in-gap states were shifted oppositely by the ISOC, and the
relative shifts of them on sublattices $A$ and $B$ were also of
opposite algebraic sign. The calculated results might serve as
experimental signatures for identifying the ISOC in the Ising
superconductors, and the scenario proposed here might also be
prospective in manipulation of electron spins in motion through the
orbital effect in the SC vortex states.

\section{acknowledgement}
\par This work was supported by the National Natural Science
Foundation of China (Grant Nos. 11574069 and 61504035) and the
Natural Science Foundation of Zhejiang Province (No. LY16A040010). This work was also supported by K. C. Wong Magna Foundation in
Ningbo University.

\end{document}